\documentclass[aps,superscriptaddress,nofootinbib,floatfix,mamsmath,11pt]{revtex4}
\usepackage{epsfig,epsf,graphics,psfrag}
\usepackage{times}
\usepackage{float}
\usepackage{color}
\usepackage{amsfonts,amssymb,stmaryrd,latexsym,amsmath,gensymb}
\usepackage{slashed}
\usepackage{bm}
\usepackage[linktocpage=true]{hyperref}
\usepackage{url}

\newcommand{\be}{\begin{equation}}
\newcommand{\ee}{\end{equation}}
\newcommand{\ba}{\begin{eqnarray}}
\newcommand{\ea}{\end{eqnarray}}

\begin{document}
\title{New structures in the proton-antiproton system}

\author{I.~T.~Lorenz}
\email{lorenzi@hiskp.uni-bonn.de}
\affiliation{Helmholtz-Institut f\"ur Strahlen- und
             Kernphysik and Bethe Center for Theoretical Physics, \\
             Universit\"at Bonn,  D--53115 Bonn, Germany}
             
\author{H.-W.~Hammer}
\email{Hans-Werner.Hammer@physik.tu-darmstadt.de}
\affiliation{Institut f\"ur Kernphysik, Technische Universit\"at Darmstadt, 64289 Darmstadt, Germany}
\affiliation{ExtreMe Matter Institute EMMI, GSI Helmholtzzentrum f\"ur
             Schwerionenforschung GmbH, 64291 Darmstadt, Germany}

\author{Ulf-G.~Mei{\ss}ner}
\email{meissner@hiskp.uni-bonn.de}
\affiliation{Helmholtz-Institut f\"ur Strahlen- und
             Kernphysik and Bethe Center for Theoretical Physics, \\
             Universit\"at Bonn,  D--53115 Bonn, Germany}
\affiliation{Institute for Advanced Simulation, Institut f\"{u}r Kernphysik
             and J\"ulich Center for Hadron Physics, \\
             Forschungszentrum J\"{u}lich, D--52425 J\"{u}lich, Germany}
             
\begin{abstract}
\noindent In the most recent measurements of the reaction $e^+e^- \rightarrow p\bar{p}$ by 
the BABAR collaboration, new structures have been found with unknown origin. We examine a 
possible relation of the most distinct peak to the recently observed $\phi(2170)$. 
Alternatively, we analyse possible explanations due to the nucleon$\,\bar{\Delta}$ and 
$\Delta\bar{\Delta}$ thresholds. The latter could explain a periodicity found in the data.
\end{abstract}

\maketitle

\section{Introduction}
\noindent 
The creation of nucleon-antinucleon $(N\bar{N})$ pairs from the electromagnetic (em) current is one of the most fundamental baryonic processes and therefore an ideal ground to examine their interactions. This process exhibits the transition between the perturbative and nonperturbative regime of QCD. Close to the production threshold, it is dominated by meson exchange that gives rise to a strong enhancement \cite{Haidenbauer:2014kja}. Increasing the center-of-mass energy, one finds an intermediate region, where this first threshold enhancement has decreased, though a description in terms of perturbative QCD is not yet possible. We will focus in particular on the region below the occurrence of the $J/\Psi$. This region is very poorly understood so far, so that a mainly conceptual analysis as performed here is appropriate.\newline
The nucleon form factors (NFFs) are the functions that parametrize the $\gamma N\bar{N}$ vertex generated by the strong interaction. In the intermediate region of interest here, an effective proton FF shows several structures. We discuss for the first time the possible main sources of these structures. On the one hand, we examine whether the newly PDG-listed vector meson $\phi(2170)$ might be relevant here. On the other hand, we investigate if such an effect could be due to the final state interaction (FSI) at the thresholds of the first resonance excitations, the $N\bar{\Delta}$ + c.c. and $\Delta\bar{\Delta}$ thresholds, respectively. A possible $N\bar{\Delta}$-cusp effect has been suggested by Rosner \cite{Rosner:2006vc}, in analogy to the case of pion photoproduction off the nucleon \cite{Bernard:1994gm}, but never examined further.\newline
For the treatment of the $\phi(2170)$, we consider a simultaneous description of the processes $e^+e^-\rightarrow N\bar{N}$ and $eN\rightarrow eN$, including observables for the proton and the neutron. Enhancements at the $N\bar{\Delta}$ + c.c. and $\Delta\bar{\Delta}$ thresholds are treated in a simplified model calculation.\newline
The unphysical region of the NFFs is equally of interest, since a $N\bar{N}$ bound state, often denoted as baryonium, would be manifest below the physical threshold. Indications for such a state around an invariant mass of $1835\,$GeV have been found in the invariant mass-spectra of the decays $J/\Psi\rightarrow xp\bar{p}$, $\Psi'(3686)\rightarrow xp\bar{p}$ with $(x = \gamma, \omega, \rho, \pi, \eta)$ and $B^+\rightarrow K^+p\bar{p}$, see e.g. Refs.~\cite{BESIII:2011aa, Aubert:2005gw, Bai:2003sw, Datta:2003iy}. The binding of a baryonium state could be generated by the final state interaction, giving rise to a pole below threshold, that could be accommodated in a recent FSI analysis \cite{Kang:2015yka}. The analytic continuation of the NFFs that we obtain from fits to data into the unphysical region is also examined with the help of logarithmic dispersion relations, including the region of a possible baryonium pole. Such an analytic continuation, however, requires the separation of the electric and magnetic NFFs over all the included kinematical range. This separation depends on the ratio between electric and magnetic FF. A higher precision for this ratio than from previous measurements is expected from the planned $p\bar{p}$-annihilation experiment PANDA at FAIR \cite{Tomasi-Gustafsson:2014pea, dbeyssi13}.\newline
For completeness, it may be worthwhile mentioning the findings of a partial wave analysis (PWA) of $pp$ elastic scattering in Ref.~\cite{Hoshizaki:1993in}, where a peak in the $^1D_2$ partial wave has been explained as due to an $S$-matrix pole at $2.144\,$GeV and related to an unstable $N\Delta$ bound state. Also a recent PWA by the SAID collaboration \cite{Arndt:2007qn} includes such $pp$ data and finds a clear signal in the $^1D_2$ amplitude around this energy. Similarly, just below the $\Delta\Delta$ threshold, a recent PWA including new neutron-proton scattering data \cite{Adlarson:2014ozl} confirmed a pole related to the $d^*(2380)$, found at COSY \cite{Bashkanov:2008ih}, suggested to be a dibaryon. This raises the question whether similar mechanisms are at play in the $p\bar{p}$ system.\newline
The paper is structured as follows. In the rest of this introductory section, we give some basic definitions, discuss the contributions to the NFFs and experimental input. In Sec.~\ref{sec:phi}, we examine a possible manifestation of the $\phi(2170)$ in different nucleon observables. The inclusion of threshold cusps is examined in Sec.~\ref{sec:thr}. In Sec.~\ref{sec:unph}, we consider the analytical continuation of the NFFs to the unphysical region. We conclude with a discussion in Sec.~\ref{sec:disc}.

\subsection{Definitions and prerequisites}
\noindent 
For the description of the em process $e^+(p_1)e^-(p_2) \to p(p_3)\bar{p}(p_4)$ we choose the center-of-mass (CM) frame, i.e. $p_{1,2} = (E,\pm k_e)$ and $p_{3,4} = (E,\pm k_p)$. The photon momentum $q$ then determines the center-of-mass energy by $q^2 = (p_1+p_2)^2 = E_{CM}^2 = (2E)^2$. In our metric timelike $q$ implies positive $q^2$. The three-momenta $k_e,k_p$ appear in the phase-space factor $\beta = k_p/k_e$,  which in the limit of neglecting the electron mass yields $\beta \approx k_p/E = \sqrt{1 - 4m_p^2/q^2}$, the velocity of the proton, and $m_p$ is the proton mass. We denote the emission angle of the proton by $\theta$. The differential cross section in the one-photon-exchange approximation in this notation is
\begin{align}
 \frac{d\sigma}{d\Omega} 
 &= \frac{\alpha^2\beta}{4 q^2}C(q^2)\left[(1+\cos^2\theta)|G_M(q^2)|^2 + \frac{4m_p^2}{q^2}\sin^2\theta|G_E(q^2)|^2\right], 
\end{align}
where $G_E$ and $G_M$ denote the electric and magnetic Sachs form factors, respectively, and $\alpha = e^2/(4\pi) = 1/137.06$ the fine-structure constant. $C(q^2)$ is the Sommerfeld-Gamow factor that accounts for the Coulomb interaction between the final-state particles
\begin{align}
 C(q^2)=\frac{y}{1-e^{-y}},\hspace{8pt} y=\frac{\pi\alpha m_p}{k_p}.
\end{align}
Integrating over the full angular distribution gives the total cross section
\begin{align}
 \sigma_{e^+e^- \rightarrow p\bar{p}}(q^2) &= \frac{4\pi\alpha^2\beta}{3q^2}C(q^2)\left[|G_M(q^2)|^2+\frac{2m_p^2}{q^2}|G_E(q^2)|^2\right]\notag\\
 &\equiv \frac{4\pi\alpha^2\beta}{3q^2}C(q^2)\left(1+\frac{2m_p^2}{q^2}\right)|G_{\rm eff}^p(q^2)|^2.
\end{align}
Thus, eliminating the kinematical factors from $\sigma$ defines the effective form factor $G_{\rm eff}$
\begin{align}
 \left|G_{\rm eff}\right| \equiv \sqrt{\frac{|G_E|^2+\frac{q^2}{2m_p^2}|G_M|^2}{1+\frac{q^2}{2m_p^2}}}.\label{geff}
\end{align}
For neutrons, the formulas are equivalent except for the Sommerfeld-Gamow factor which is not present in that case. Beyond the Coulomb FSI, higher order QED corrections will be neglected in this work. The weak neutral current contribution to the measured cross section is also neglected. For the time-reversed process, the phase space factor is inverted, yielding $\sigma_{e^+e^- \to p\bar{p}} = \beta^2\sigma_{p\bar{p} \to e^+e^-}$.\newline
Taking into account the angular dependence of $p\bar{p}$ production, one can express the differential cross section via the angular asymmetry $\mathcal{A}$,
\begin{align}
 \frac{d\sigma}{d\Omega} = \left.\frac{d\sigma}{d\Omega}\right|_{\theta=90\degree}
 [1+\mathcal{A}\cos^2\theta],
\end{align}
with 
\begin{align}
 \mathcal{A} = \frac{\frac{q^2}{(4m_p^2)} - R^2}{\frac{q^2}{(4m_p^2)} + R^2},
\end{align}
and determine from this explicitly the FF ratio $R = |G_E/G_M|$.\newline 
For many aspects, it is instructive to consider the vertex $\gamma p\bar{p}$ in the helicity basis, i.e. the helicity-conserving Dirac and helicity-changing Pauli form factors $F_1$ and $F_2$, in order:
\begin{align}
 G_E(q^2) &= F_1(q^2) + \frac{q^2}{4m_p^2}F_2(q^2),\nonumber\\
G_M(q^2) &= F_1(q^2) + F_2(q^2).
\end{align}
On the one hand, this basis allows us to see directly that the threshold relation is by definition
\begin{align}
 G_E(4m_p^2) = G_M(4m_p^2).
\end{align}
In addition, the asymptotic $q^2$-dependence can also be conveniently given. For large 
$-q^2=Q^2 \geq 0$, the Dirac and Pauli form factors can be predicted from perturbative QCD \cite{Lepage:1979za, Lepage:1980fj} to behave like
\begin{align}
 \lim\limits_{Q^2 \to \infty} F_i(Q^2) = (Q^2)^{-(i+1)}\left[\ln\left(\frac{Q^2}{\Lambda^2_{\rm QCD}}\right)\right]^{-\gamma},\hspace{5pt}i=1,2 ~,
\end{align}
with
\begin{align}
 \gamma = 2 + \frac{4}{3\beta}\hspace{5pt}\text{and}\hspace{5pt}\beta=11-\frac{2}{3}N_f. 
\end{align}
Here, $\beta$ is the QCD $\beta$-function to one loop for the number of flavors $N_f$. The anomalous dimension $\gamma$ depends weakly on the latter, $\gamma = 2.148, 2.160, 2.173$ for $N_f = 3, 4, 5$, respectively. The analytic continuation of the logarithm to timelike momenta yields an additional term $\ln(Q^2/\Lambda^2) = \ln(q^2/\Lambda^2) - i\pi$, for $q^2 > \Lambda^2$.\newline
For the asymptotic behavior of the form factors, we consider the Phragm\'{e}n-Lindeloef theorem \cite{Hohler83}: ``Let $f(z)$ be an analytic function of $z$, regular and bounded in Im $z > 0$. If $f(z)$ tends to the limits $L_1$ and $L_2$ along the rays $z=x+i0$ as $x\to \pm\infty$, then we must have $L_1=L_2$.'' In particular, from this it follows that the imaginary part has to vanish in the asymptotic limit.\newline
Based on the reasonings of perturbative QCD including the analytic continuation into the timelike region \cite{Shirkov:1997wi, Bakulev:2000uh}, a recent analysis \cite{Bianconi:2015owa} of the proton effective FF includes a fit of the form 
\begin{align}
 |G_{\rm eff}^p(q^2)| = \frac{A}{(q^2)^2(\ln^2(q^2/\Lambda^2)+\pi^2)},\label{eq:pqcd}
\end{align}
with the parameters from a fit to data prior to a recent measurement by the BABAR collaboration \cite{Lees:2013ebn}, given as $A=72\,$GeV$^{-4}$ and $\Lambda=0.52\,$GeV.

\subsection{Possible contributions}\label{sec:cont}
\noindent 
\begin{figure}[ht]
\centering
\includegraphics[width=1\textwidth]{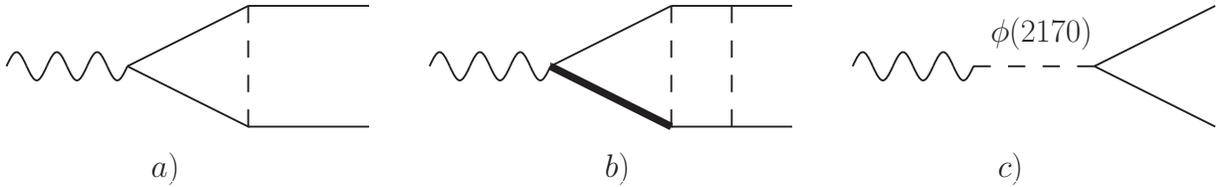}
\caption{The different types of diagrams contributing to the NFFs above the $p\bar{p}$ threshold, as discussed in Sec.~\ref{sec:cont}. The wiggly line denotes the photon, the thin solid line the (anti)nucleon, the thick solid line an (anti)nucleon resonance and the dashed line denotes all possible mesons, e.g. pions in a)+b).}\label{fig:tff}
\end{figure}
In the energy-regime below the $J/\Psi$, we will focus on possible hadronic contributions to the NFFs. The size of the different possible contributions is unknown. Therefore we consider here the individual diagrams in a pioneering study, neglecting interference effects between them.\newline
Three Feynman graphs representing the different types of diagrams contributing to the NFFs are shown in Fig.~\ref{fig:tff}. The first two refer to a mainly baryonic, the last to a mesonic contribution. Diagram $a)$ represents the final-state interaction (FSI) in form of meson exchange. The  pion exchange shown  can be replaced by any number of suitable mesons. Diagram $b)$ shows one possible excitation of a resonance, e.g. a $\Delta$, in the FSI diagram. Possible re-excitations are assumed. Diagrams of type $c)$ are usually ignored above the $p\bar{p}$ threshold. However, in general they can contribute, e.g. from the $\phi(2170)$. 

\subsection{Data}\label{sec:data}
\noindent 
\begin{figure}[ht]
\centering
\includegraphics[width=0.6\textwidth]{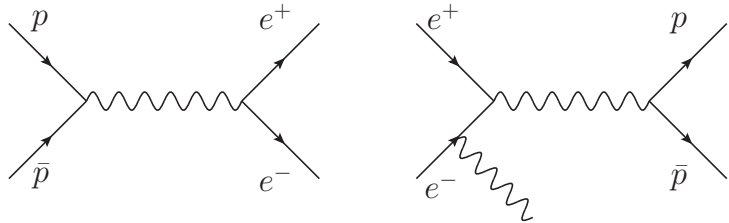}
\caption{The processes mainly considered for NFF determinations. Left: Annihilation measured at LEAR, planned at FAIR \cite{Sudol:2009vc, Tomasi-Gustafsson:2014pea, dbeyssi13}. Right: Production with ISR, measured e.g.  at BABAR.}\label{fig:timeff}
\end{figure}
\begin{figure}[ht]
\centering
\includegraphics[width=0.6\textwidth]{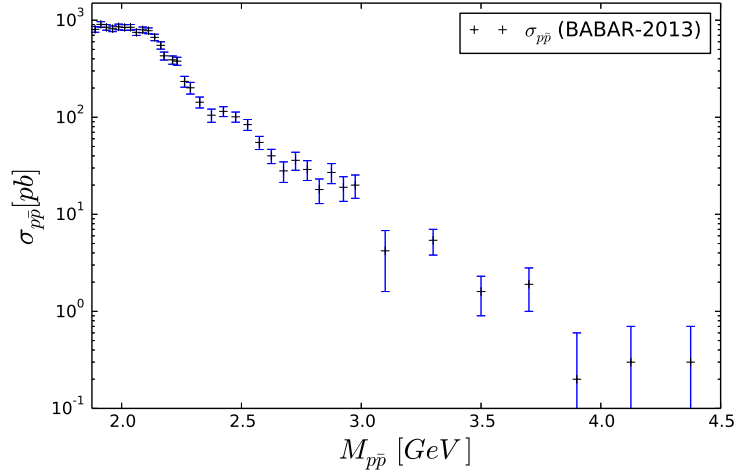}
\caption{The total cross section $\sigma(e^+e^-\to p\bar{p})$ measured at BABAR, depending on the invariant mass $M_{p\bar{p}}$. The initial state radiation is factorized out and the $J/\Psi$ and $\Psi(2s)$ peaks have been removed \cite{Lees:2013ebn}.}\label{fig:illcross}
\end{figure}
\begin{figure}[ht]
\centering
\includegraphics[width=0.6\textwidth]{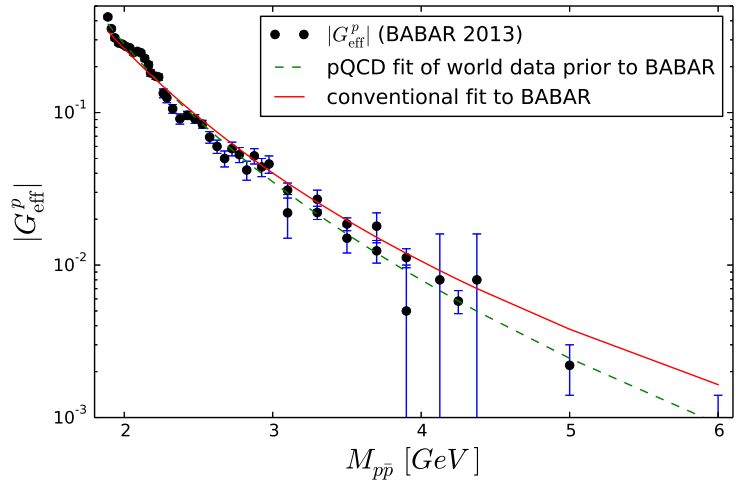}
\caption{The effective form factor $G_\text{{eff}}^p$ of the proton measured in the process $\sigma(e^+e^-\to p\bar{p}\gamma)$ at BABAR \cite{Lees:2013ebn}, shown with different fits from Ref.~\cite{Bianconi:2015owa}, given by a conventional dipole$\cdot$monopole form or the pQCD parametrization Eq.~\eqref{eq:pqcd}.}\label{fig:illgeff}
\end{figure}
Experimental information on the proton FFs for timelike momenta is available from a number of measurements dating back to 1976, for a detailed discussion, see Ref.~\cite{Denig:2012by}.  However, the only ones that include sufficient precision on the angular distribution to disentangle $G_E$ and $G_M$, are those from LEAR \cite{Bardin:1994am}, BABAR \cite{Lees:2013ebn}, see Fig.~\ref{fig:timeff}, and BESIII \cite{Ablikim:2015vga}.  In fact, the first dispersion analyses of timelike nucleon form factor data~\cite{Hammer:1996kx} were hampered by a missing separation of $G_E$ and $G_M$.
The recent BESIII measurement yields good agreement with BABAR, in particular confirming an enhancement at $2.4$ GeV. However, the emphasis of BESIII lies above the region of our main interest here. Due to the much higher precision of the BABAR data compared to the ones from LEAR, in this work we focus on the former, for the corresponding cross sections see Fig.~\ref{fig:illcross}. The ridge and bump structure occurring here are also inherited by the effective proton FF that is shown in Fig.~\ref{fig:illgeff}. The insufficiency of a pQCD description is shown by the fit Eq.~\eqref{eq:pqcd} from Ref.~\cite{Bianconi:2015owa}, where it is argued that the oscillations around the pQCD fit are due to FSI.\newline
In this work, the angular information is included in the form of the FF ratio $|G_E/G_M|$ as provided by BABAR \cite{Lees:2013ebn}.\newline
The neutron FFs for timelike momenta have been measured in the process $e^+e^-\rightarrow n\bar{n}$ by the FENICE collaboration \cite{Antonelli:1998fv} and recently at the VEPP-2000 collider with the ``spherical neutral detector'' (SND) \cite{Achasov:2014ncd}. As for the proton case, the precision of the latest neutron measurement exceeds by far that of the preceding experiments. However, the accuracy does not suffice to determine the neutron form factor ratio.\newline
In the spacelike region, we include explicitly the most precise differential $ep$ scattering cross sections \cite{Bernauer10} and in addition the polarization world data (see Ref.~\cite{Punjabi:2015bba} for a list). 
For the neutron, we want to emphasize that the phenomenological extraction of neutron FFs yields larger uncertainties than for the proton case, since there is no free target. The measurements require light nuclei targets like $^2$H or $^3$He for quasi-elastic scattering, for details see e.g. Ref.~\cite{Punjabi:2015bba}. Here, we use the data on $G_M^n$ and $G_E^n$ for a better visibility compared to the equivalent use of $G_M^n$ and the ratio.

\section{The process $e^+e^- \to N\bar{N}$ and the $\phi(2170)$}\label{sec:phi}
\noindent 
In this section we consider the possible contribution to the NFFs from the $\phi(2170)$, corresponding to diagram $c)$ in Fig.~\ref{fig:tff}. This refers to a structure that has been found in different processes and that is at the moment classified by the PDG as the only light unflavored vector meson above the $N\bar{N}$-threshold. In the following, we use the PDG notation even though some of these structures have also been denoted as $X/Y(2175)$ or simply as $\phi''$. As shown in Tab.~\ref{table:phi}, measurements at BES, BABAR and BELLE have found signals in this mass region, albeit over some interval. Even the central values spread in the range $2.08-2.19$~GeV for the mass and $58-192$~MeV for the width. This might correspond to the uncertainty of the separation from non-resonant background and/or the possible existence of multiple interfering resonances in this range. Also the isospin is given as definite, yielding altogether $I^G(J^{PC}) = 0^-1^{--}$. However, the assignment of quantum numbers should be taken with a grain of salt, mainly due to the limited statistics, for a discussion see for example Ref.~\cite{Ablikim:2010au}. Different suggestions about the origin of the $\phi(2170)$ have been put forward. It has been interpreted, for example, as a tetraquark state \cite{Ali:2011qi}, a hybrid $s\bar{s}g$ resonance \cite{Ding:2006ya} or to a large extent as a $\phi(1020)K\bar{K}$ state \cite{MartinezTorres:2008gy}. It can also be generated in a chiral Lagrangian approach for $\phi(1020)/f_0(980)$ $S$-wave scattering by their self-interactions \cite{AlvarezRuso:2009xn}. However, to our knowledge it has not been considered in relation to the NFFs. We will do so, first by focusing on the effective proton FF and second in a simultaneous treatment of different proton and neutron measurements for space- and timelike momenta. The individual form factors that correspond to a Breit-Wigner resonance structure with mass $M_{\phi}$ and width $\Gamma_{\phi}$ behave like $F(q^2) \propto 1/(M_{\phi}^2-q^2-i\Gamma_{\phi}M_{\phi})$, so that the effective FF can be fitted to the absolute value of the latter. 
\begin{table}[ht!]
 \centering
\begin{tabular}{|l|c|c|}
\hline
process & mass (MeV) &  width (MeV) \\
\hline
$J/\Psi\to\eta\phi f_0(980)$ [BES]& $2186\pm10\pm6$ & $65\pm23\pm17$ \\
$e^+e^-\to \phi\eta\gamma$ [BABAR]& $2125\pm22\pm10$ & $61\pm50\pm13$ \\
$e^+e^-\to K^+K^-\pi\pi\gamma$ [BABAR]& $2175\pm10\pm15$ & $58\pm16\pm20$ \\
$e^+e^-\to K^+K^-\pi^+\pi^-\gamma$ [BELLE]& $2079\pm13^{+79}_{-28}$ & $192\pm23^{+25}_{-61}$ \\
$e^+e^-\to K^+K^-\pi^+\pi^-\gamma$ [BABAR]& $2192\pm14$ & $71\pm21$ \\
$e^+e^-\to K^+K^-\pi^0\pi^0\gamma$ [BABAR]& $2169\pm20$ & $102\pm27$ \\
\hline
\end{tabular}
\caption{Observations of the $\phi(2170)$ from the review of particle properties~\cite{Agashe:2014kda}.}
\label{table:phi}
\end{table}

\subsection{Individual $G_{\rm eff}^p$ fits}\label{sec:gbw}
\noindent 
A description of the proton effective form factor has been attempted by several groups over the years with the  main emphasis on either the perturbative QCD part or the impact of vector mesons below the $p\bar{p}$ threshold. Neither these nor conventional fits of dipoles, monopoles or products of these can fully accommodate the structures in the currently most precise relevant cross sections or in $G_{\rm eff}^p$, see Figs.~\ref{fig:illcross}, \ref{fig:illgeff}.  
\begin{figure}[ht]
\centering
\includegraphics[width=0.5\textwidth]{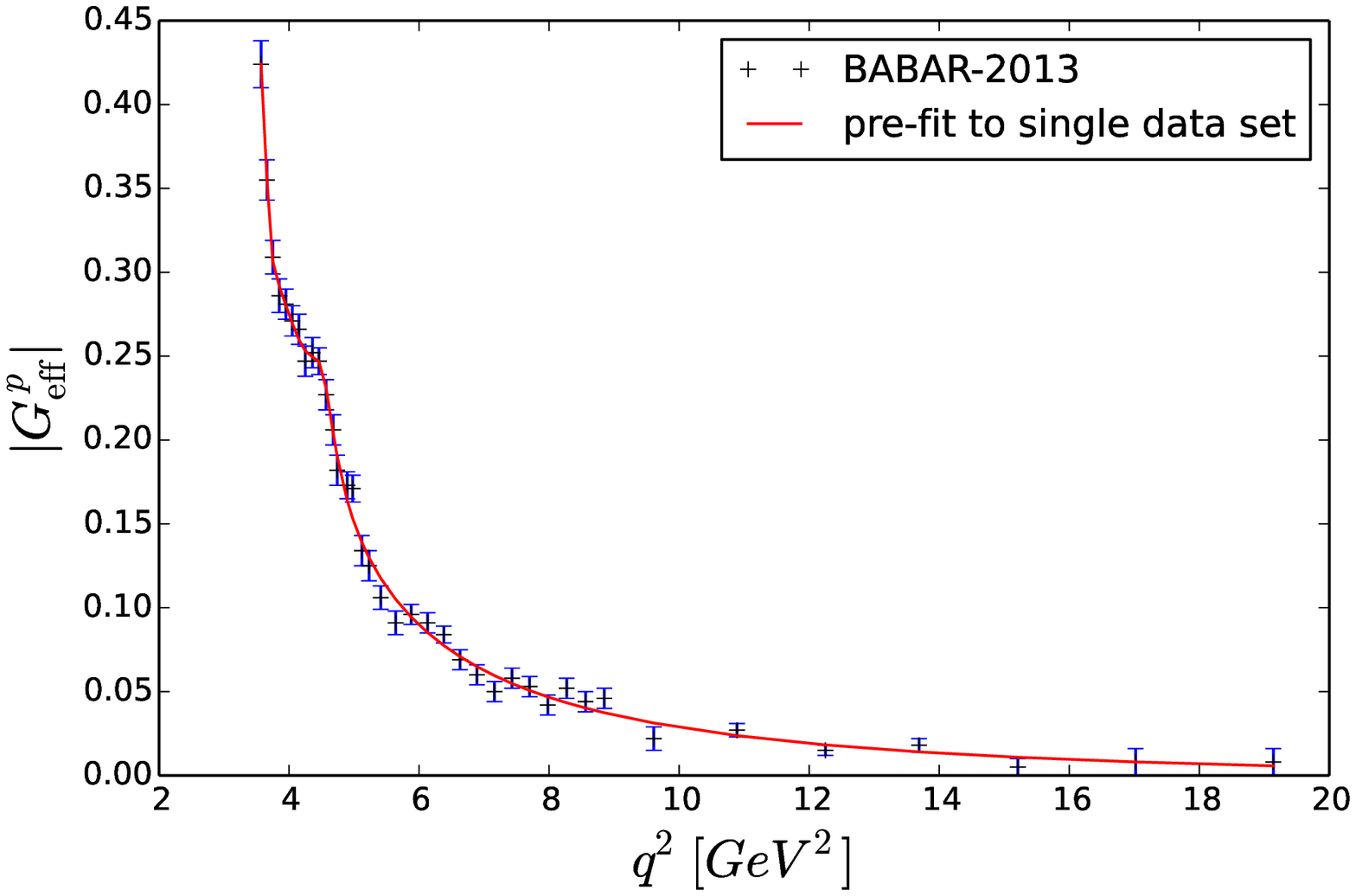}\hglue1mm
\includegraphics[width=0.5\textwidth]{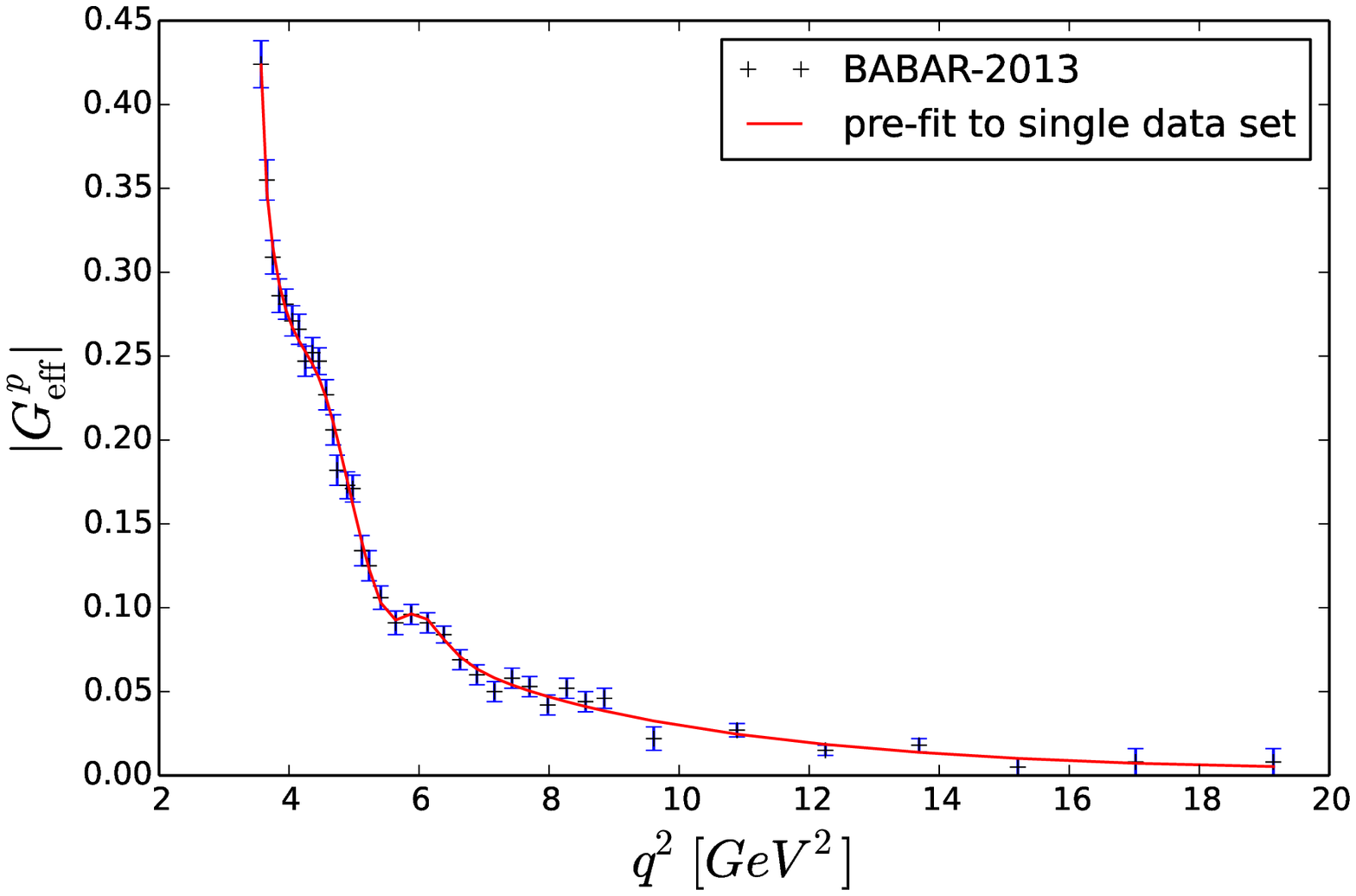}
\caption{Illustrative pre-fits for comparison to PDG-given $\phi(2170)$, using only $G_{\rm eff}$-data.  Including a resonance at $M_1 = 2.125$~GeV with a width of $\Gamma_1 = 0.09$~GeV (left) and $\Gamma_1 = 0.33$~GeV (right). The fit on the right also includes a resonance at $M_2 = 2.43$~GeV and $\Gamma_2 = 0.23$~GeV.}\label{fig:bw}
\end{figure}
In the overall concave function $G_{\rm eff}^p$, the first structure occurs as a mainly convex part for invariant masses of around $2-2.25$~GeV. A satisfactory fit to the data requires to take this structure into account in some way in the parametrization. However, this is a prime example of the ambiguities that can appear in separating a possible resonance structure from the background. For comparison with PDG-values, see Tab.~\ref{table:phi}, we perform test fits to $G_{\rm eff}^p$ with terms that correspond to Breit-Wigner (BW) shapes in the cross section and 5 effective pole terms below threshold for the background description. We find that a large range of values of masses and widths in this region strongly improves the data description, $M \approx (2 - 2.18)$~GeV and $\Gamma\approx (0.05 - 0.5)$~GeV. In Fig.~\ref{fig:bw}, we show examplary  fits with a narrow resonance on the left and a broader one on the right. Remarkable is also a structure peaked around $q^2\approx 5.9$~GeV$^2$. Allowing a second resonance in this region yields one at $M = 2.43$~GeV with a width of $\Gamma = 0.23$~GeV. This cannot be regarded as a rigorous analysis, since these resonance structures largely overlap and are not separable from the background either. However, it is undebatable that an additional structure peaked around $M \approx 2.43$~GeV improves the data description. 

\subsection{Simultaneous fits}
\noindent 
In this section, we combine the $G_{\rm eff}^p$ fits from the last section with more available data on the NFFs, see Sec.~\ref{sec:data} for references. These data comprise 7 different sets, 4 for the proton and 3 for the neutron. For the proton, we consider the differential cross sections and the ratio $G_E/G_M$ from polarization observables on the scattering side in addition to the effective FF and $|G_E/G_M|$ on the production side. For the neutron, we include $G_E$ and $G_M$ from scattering data and again the effective FF on the production side. In order to weight the different data sets equally, their impact on the $\chi^2$-function to be minimized is determined by their number of data points. Still, to avoid a dominance by the other sets, we fix the mass of the resonance structure at $M = 2.125$~GeV. The width obtained in the simultaneous fit is $\Gamma = 0.088$~GeV. The larger number of data points compared to the previous section requires more effective pole terms in the unphysical region, in particular for a separation of the isospin channels due to the inclusion of the neutron. As the basic framework, we proceed in a similar way to our analysis of spacelike NFFs \cite{Lorenz:2014yda}. To be specific, we include parametrizations of the $2\pi$, $K\bar{K}$ and $\rho\pi$-continuum, the $\omega$- and $\phi$-contribution and effective pole terms. Here, the latter are restricted to the region $1$~GeV$^2 < q^2 < 3.52$~GeV$^2$ and limited to a number of three in the isoscalar and five in the isovector channel.\newline
\begin{figure}[ht]
\centering
\includegraphics[width=0.5\textwidth]{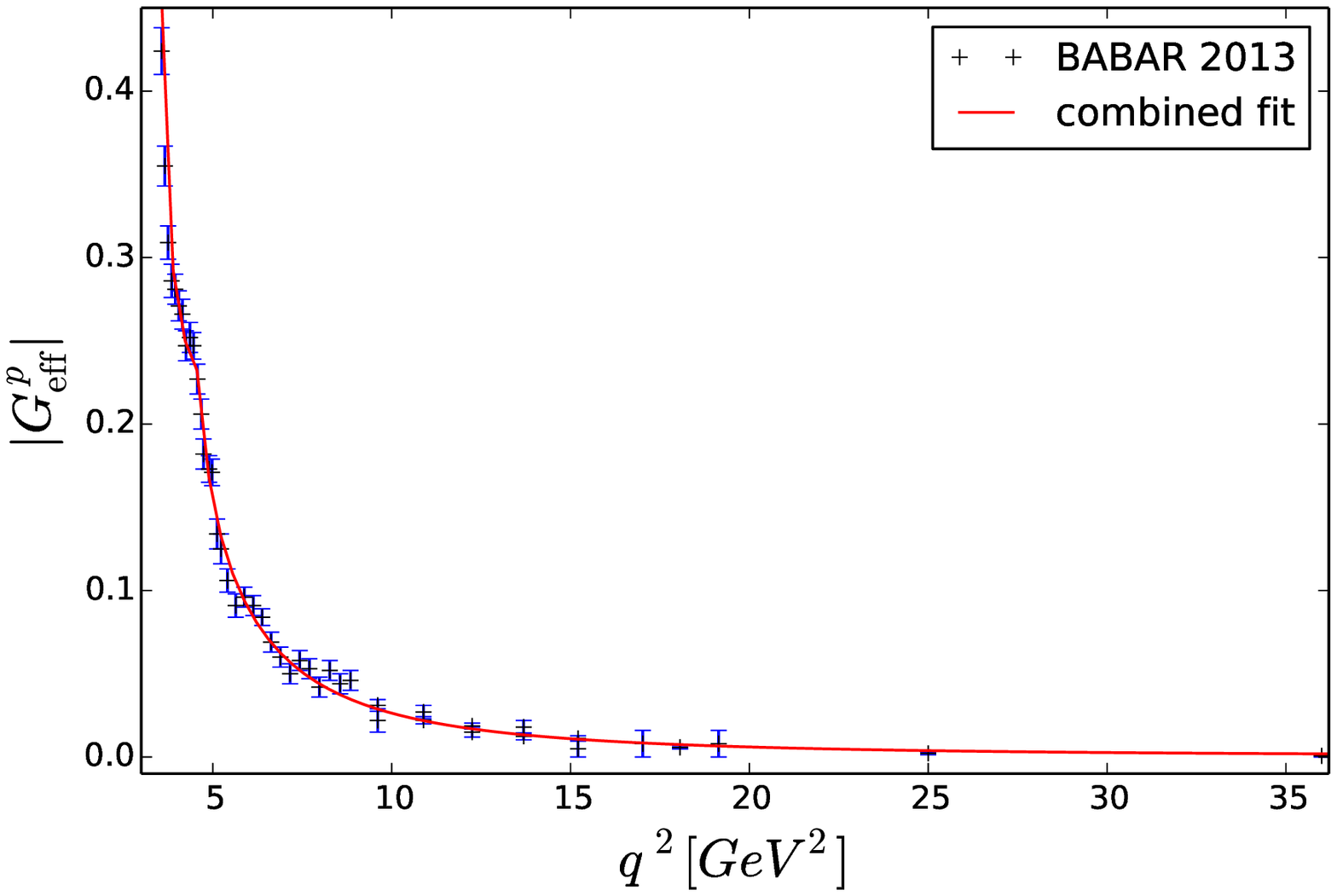}\hglue1mm
\includegraphics[width=0.5\textwidth]{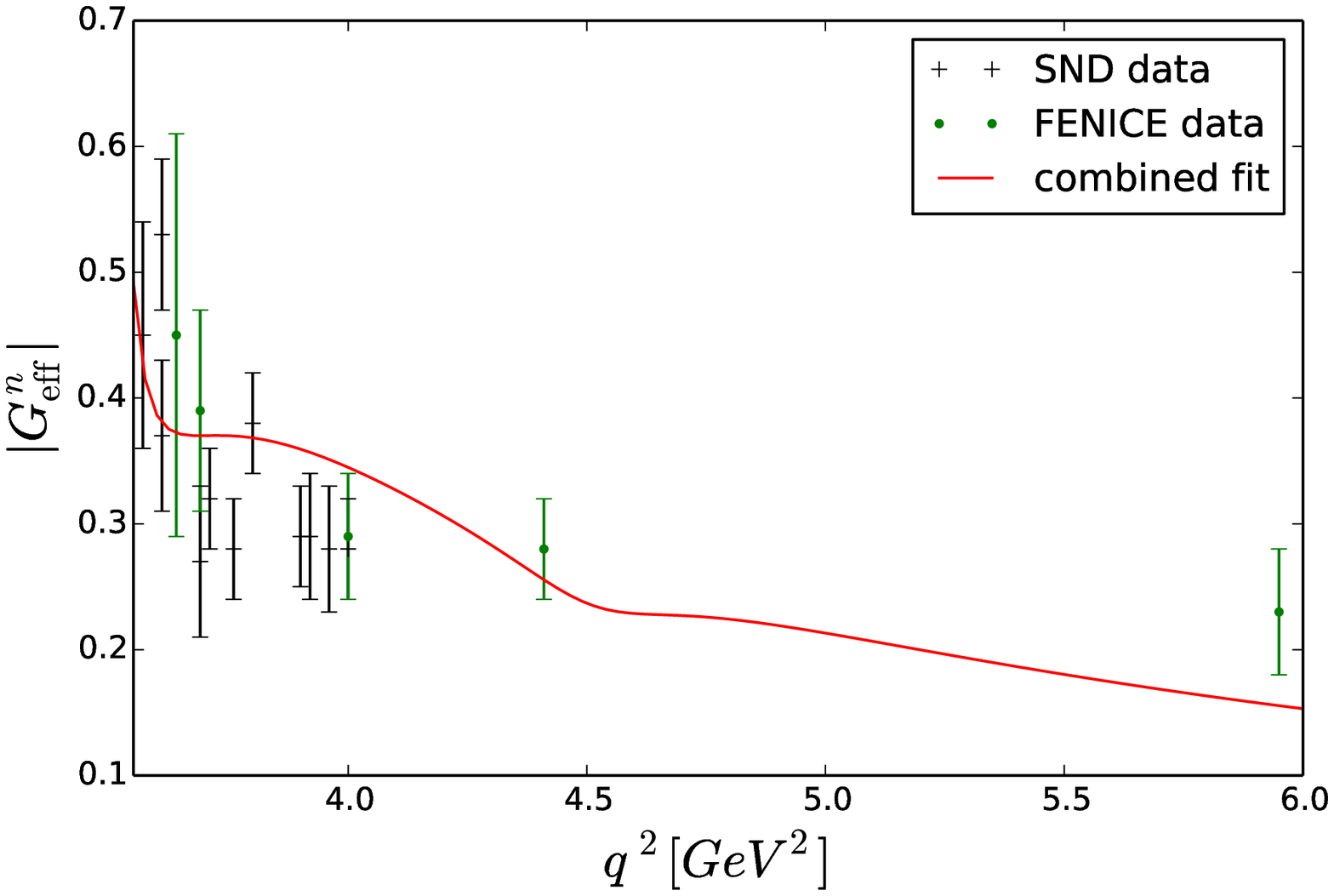}
\caption{The effective FF of the proton (left) and the neutron (right). $G_{\rm eff}^p$ is complemented here by the subsequently published higher-energy data \cite{Lees:2013uta}. The recent SND measurement \cite{Achasov:2014ncd} is in good agreement with the only previously published $G_{\rm eff}^n$ data from FENICE \cite{Antonelli:1998fv}.}\label{fig:eff}
\end{figure}
\begin{figure}[ht]
\centering
\includegraphics[width=0.5\textwidth]{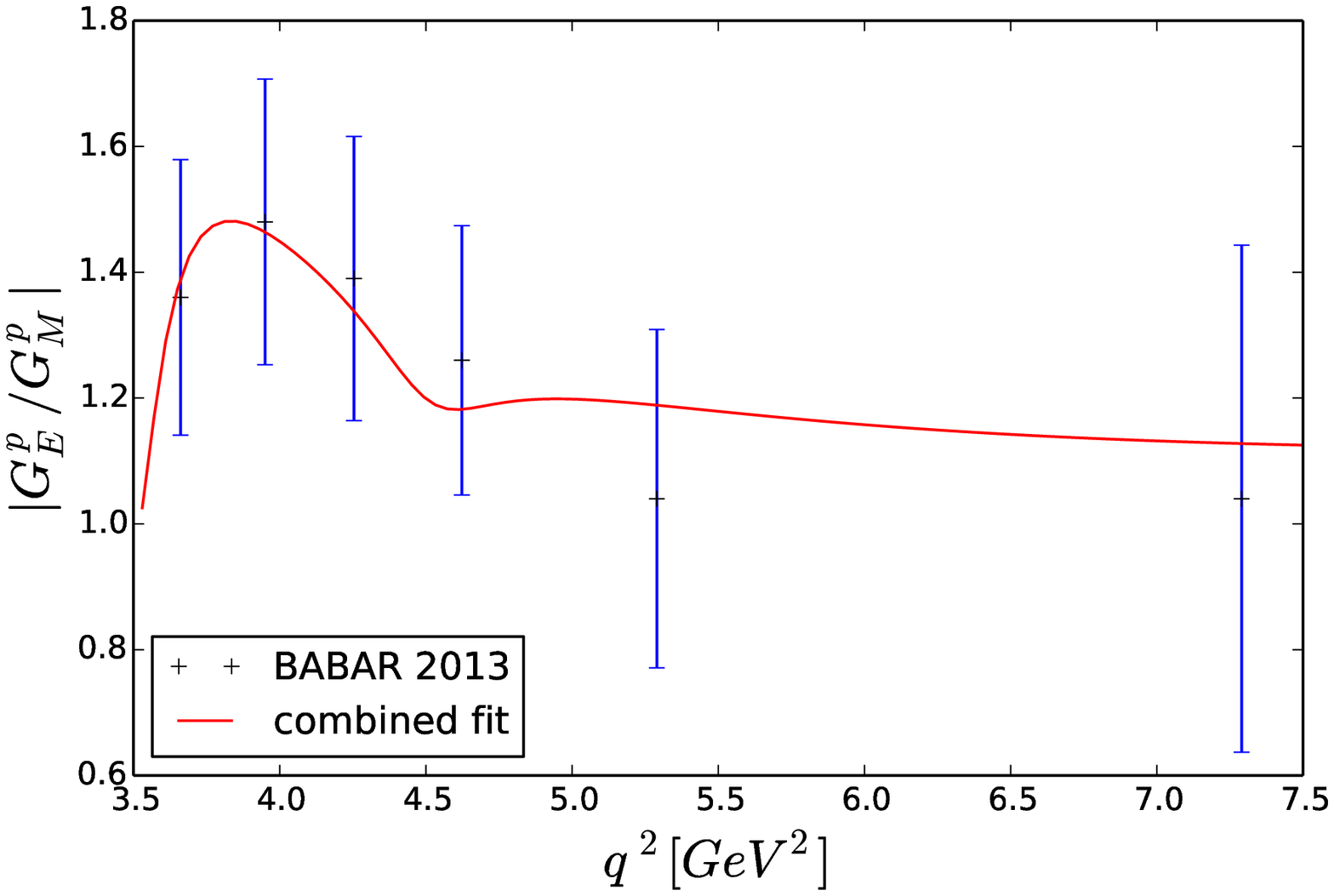}\hglue1mm
\includegraphics[width=0.5\textwidth]{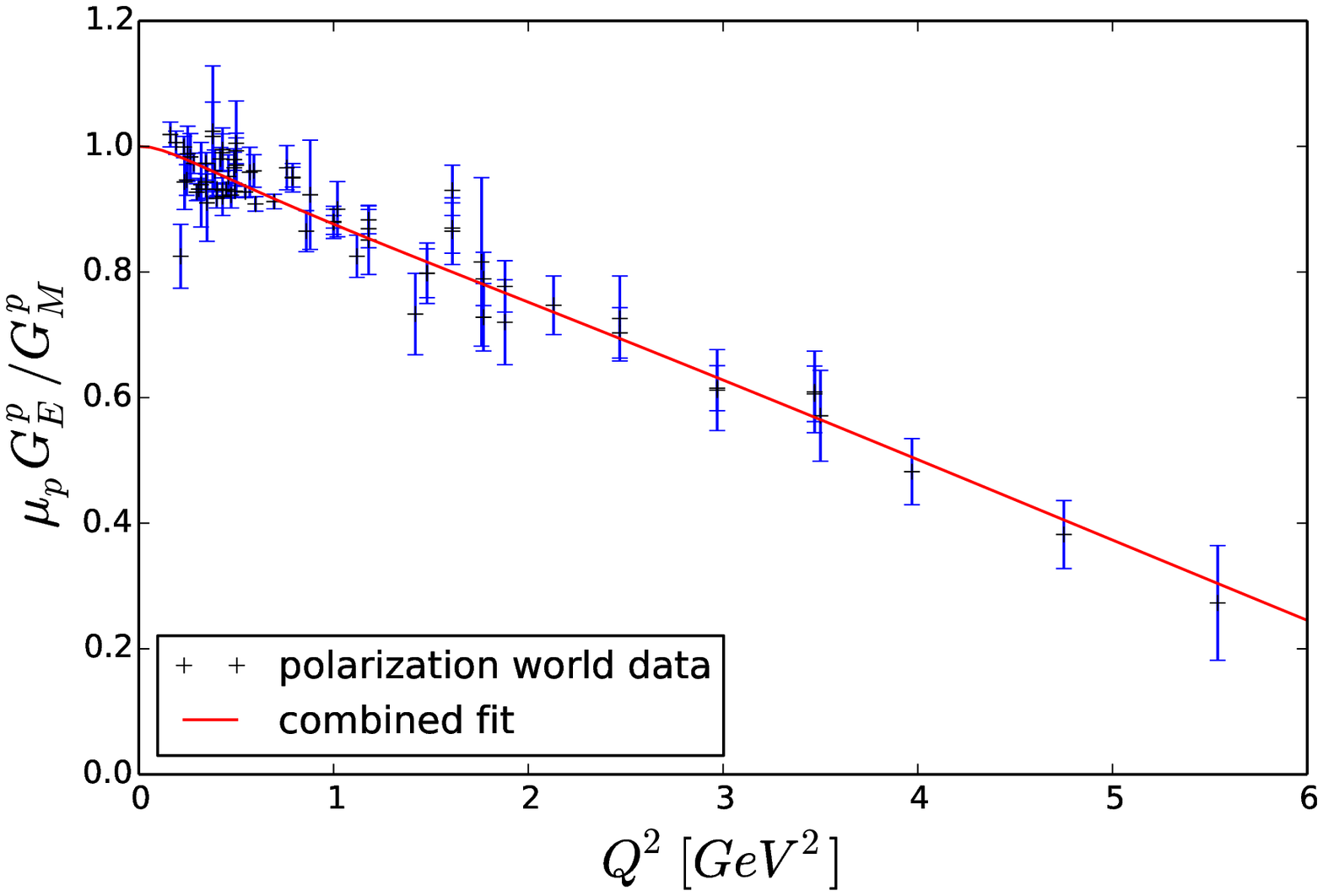}
\caption{The form factor ratio of the proton for space- (left) and timelike (right) momenta from the combined fit to space- and
timelike data.}\label{fig:ratio}
\end{figure}
\begin{figure}[ht]
\centering
\includegraphics[width=0.5\textwidth]{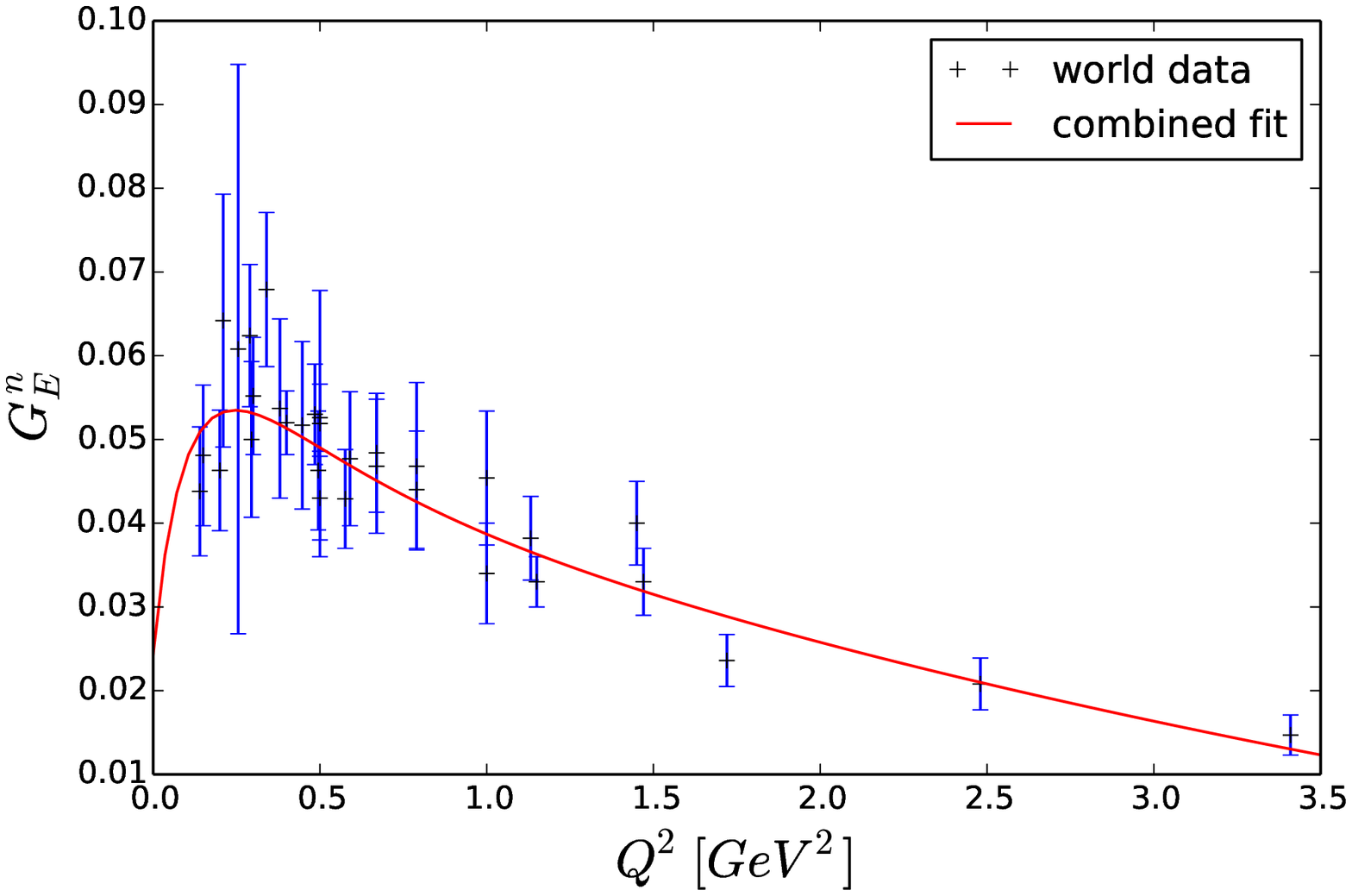}\hglue1mm
\includegraphics[width=0.5\textwidth]{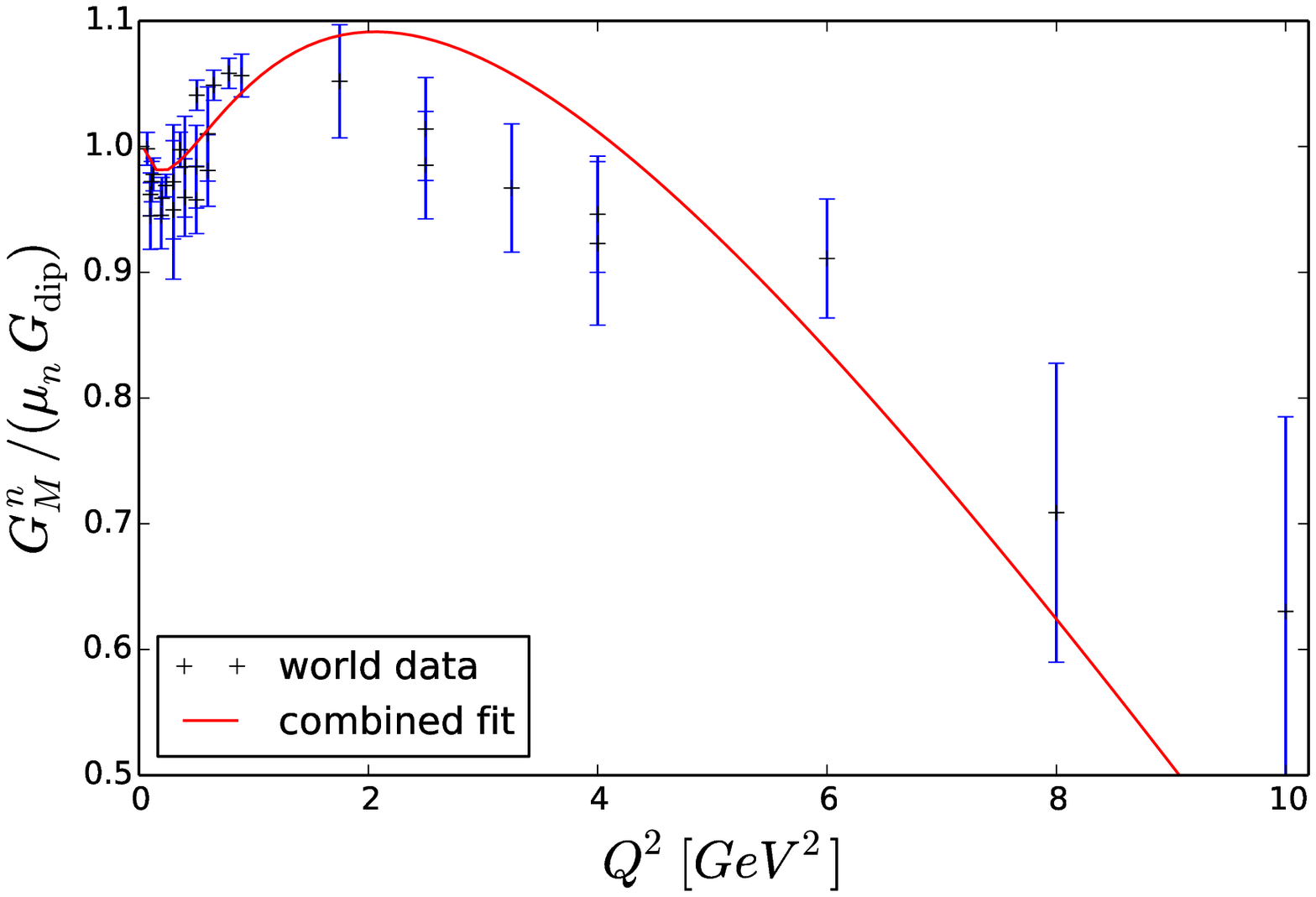}
\caption{The neutron electric (left) and magnetic (right) form factor from the combined fit to space- and
timelike data.}\label{fig:gn}
\end{figure}
\begin{figure}[ht]
\centering
\includegraphics[width=0.6\textwidth]{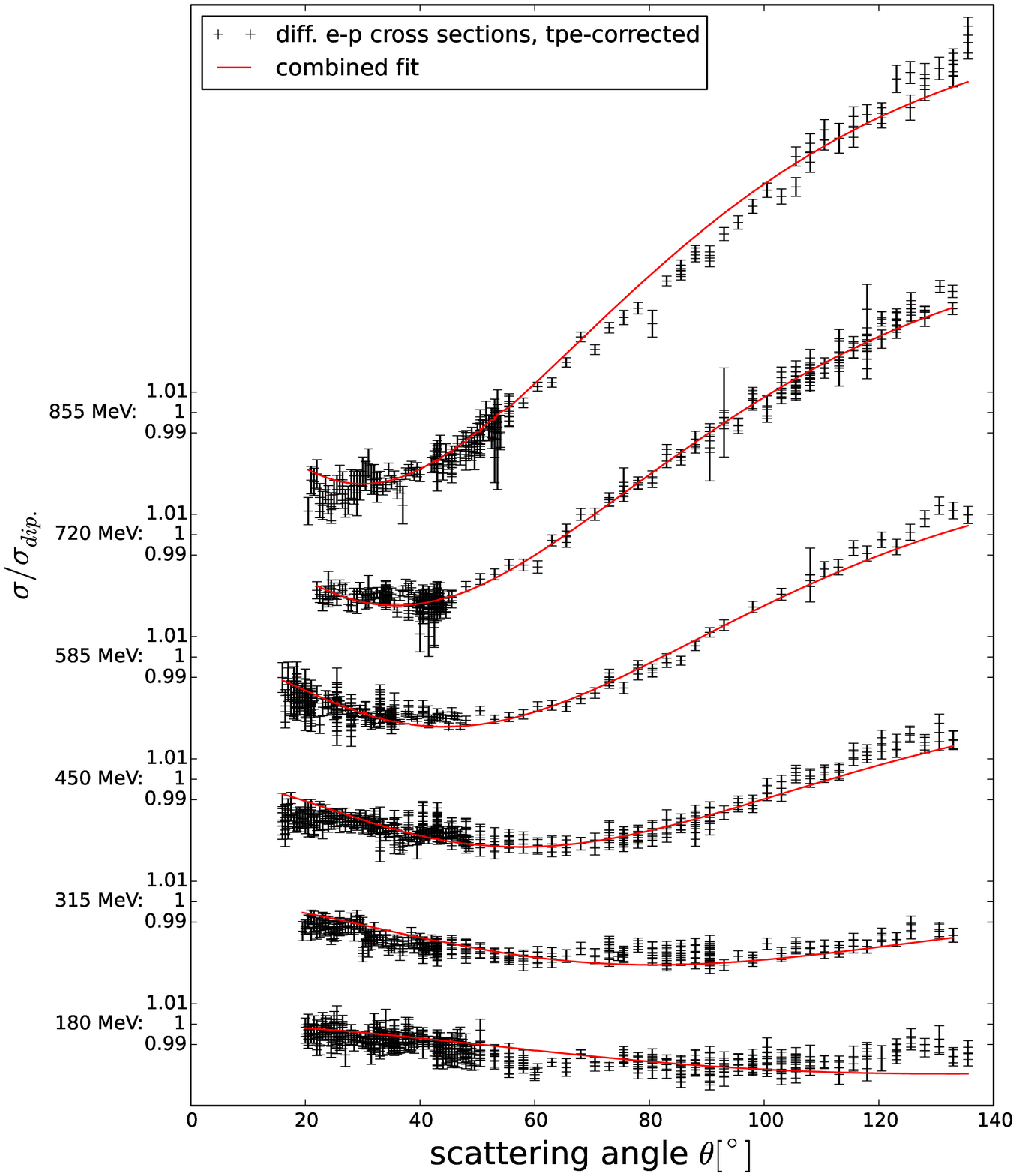}
\caption{The electron-proton scattering cross sections from MAMI \cite{Bernauer10}. On the left, the energy of the incoming electron is given, which together with the scattering angle determines $Q^2$.}\label{fig:cross}
\end{figure}
We show the fit results in Figs.~\ref{fig:eff},~\ref{fig:ratio},~\ref{fig:gn},~\ref{fig:cross}. Note that the first three subgraphs refer to the timelike and the remaining four to the spacelike region. For the proton effective FF, we included a second set of BABAR data at higher energies, that are still well described by our fit. For the neutron we included the recent measurement from the VEPP-2000 $e^+e^-$ collider \cite{Achasov:2014ncd}. In the range from threshold up to $q^2 = 4$~GeV$^2$, their $q^2$-dependence is very similar to the proton case. Only above this, two further data points given by the FENICE collaboration indicate a less steep fall-off. Unfortunately, the neutron data is too sparse in this region to constrain a possible manifestation of the $\phi(2170)$ around $q^2 = (2.125\,{\rm GeV})^2 \approx 4.5$~GeV$^2$. Also, the current level of statistics of the angular distribution of $n\bar{n}$ events is too low to determine the $|G_E^n/G_M^n|$ ratio \cite{Achasov:2014ncd}.\newline
So we are left with the $|G_E^p/G_M^p|$ ratio to search for direct indications of a resonance at $q^2 \approx 4.5$~GeV$^2$. Indeed, a slight dip occurs here, see Fig.~\ref{fig:ratio}, as soon as we include the resonance term. The same quantity in the spacelike region is also well reproduced. For the neutron FFs in the spacelike region, see Fig.~\ref{fig:gn}, we want to emphasize the sizeable uncertainties in their extraction from electron scattering off light nuclear targets like $^2H$ or $^3He$, for details see e.g. Ref.~\cite{Punjabi:2015bba}. Finally, the electron-proton scattering cross sections shown in Fig.~\ref{fig:cross} are by far the largest data set with 1422 out of the total 1627 points. Therefore weighting each set equally disfavors the larger sets in a sense, still giving reasonable agreement in this case. One could also here refit the normalization individually for different parts and treat the uncertainties as discussed in detail in Ref.~\cite{Lorenz:2014yda}. However, for a more conceptual work like the one at hand we refrain from such a procedure.

\section{Threshold enhancements}\label{sec:thr}
\noindent 
In this part, we outline an alternative origin of the structures found in the $G_{\rm eff}^p$ data by carrying out fits to $G_{\rm eff}^p$ only.
Remarkable in the last section is the position of the peaks, or rather kinks, that are necessary to improve the fits. The positions can be chosen as the threshold energies of the $p\bar{\Delta}$ + c.c. and the $\Delta\bar{\Delta}$ states, or $p\bar{p}2\pi$ and $p\bar{p}4\pi$, respectively. The occuring $\Delta$ resonance would emit a pion $(>99\%)$, or a photon $(<1\%)$. Since the backgrounds to the $p\bar{p}$ final states are subtracted, as discussed in detail by the BABAR collaboration \cite{Lees:2013ebn}, the pion or photon has to be absorbed by the other (anti)baryon. In general, the interaction between the final states can of course comprise further exchange of on- or off-shell mesons which should be treated systematically in an effective field theory (EFT) framework. Close to the $p\bar{p}$ threshold, the final-state interaction (FSI) can be computed via a Lippmann-Schwinger equation. Such a procedure has recently been updated in Ref.~\cite{Haidenbauer:2014kja}, employing a static interaction potential derived in chiral EFT \cite{Kang:2013uia} that has been fitted to a partial wave analysis of $p\bar{p}$ scattering data \cite{Zhou:2012ui}. Moreover, this is based on the assumption of a real and constant bare vertex function. In the region of  validity of such an approach $(M_{p\bar{p}} - 2m_p) \leq 0.1$~GeV, the decrease in $G_{\rm eff}^p$ can be well reproduced. Beyond this range, a calculation in this framework breaks down. However, one naively expects the interaction to further decrease due to the increasing relative velocity of the two final states. In such a naive reasoning, the excitation of a resonance would lead to the same threshold kinematics, just shifted in energy, and thus could give rise to an enhanced FSI. After the decay of the $\Delta$ resonance, which dominates here, one would be left with mainly the $p\bar{p}$ FSI. For the possible necessity to resum the loops corresponding to re-excitations, future work could proceed along the lines of Ref.~\cite{Guo:2014iya}. This might allow us to distinguish the possible origins of the structures found in $G_{\rm eff}^p$. For the moment, however, we merely illustrate some possible contributions from the triangle diagram shown in Fig.~\ref{fig:triangle}. Also these are only roughly approximated due to the lack of information on the vertices, in particular their momentum dependence.

\subsection{Inclusion of the $N\bar{\Delta}$ + c.c. and $\Delta\bar{\Delta}$ thresholds}\label{sec:model}
\noindent 
\begin{figure}[ht]
\centering
\includegraphics[width=0.3\textwidth]{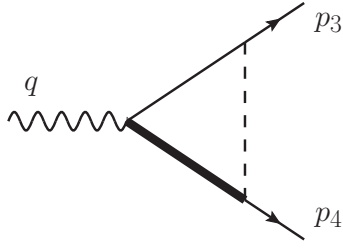}
\caption{Triangle graph with virtual $N\bar{\Delta}\pi$ state. Notations are as in Fig.~\ref{fig:tff}.}\label{fig:triangle}
\end{figure}
We consider the triangle graphs with virtual $N\bar{\Delta}\pi$, see Fig.~\ref{fig:triangle}, and $\Delta\bar{\Delta}\pi$ in order to approximate possible cusp effects. However, the vertices are not well known for these kinematics. While, e.g., for the $\Delta N\pi$ transition the coupling constant at $Q^2 = -M_{\pi}^2$ are known, the form factors and their general dependence on the different momenta is all but well known. This would be relevant if we were to evaluate the triangle diagram in full glory, which we do not attempt here.
What can be obtained most easily though, is the scalar part of the integral. This is proportional to the analytically well-known Passarino-Veltman integral $C_0(\kappa)$. As defined in Ref.~\cite{'tHooft:1978xw}, (slightly different conventions), and implemented in LoopTools \cite{Hahn:1998yk}, this depends on the configuration $(\kappa)$ of virtual particle masses and external particle four-momenta:
\begin{align}
 C_0(\kappa = p_k, p_l, m_1, m_2, m_3) = \frac{1}{i\pi^2}\int \frac{d^4k}{[k^2 - m_1^2][(k-p_k)^2 - m_2^2][(k+p_l)^2 - m_3^2]}
\end{align}
with $\kappa_1 = p_3, p_4, M_{\pi}, m_N, m_{\Delta}$ and $\kappa_2 = p_3, p_4, M_{\pi}, m_{\Delta}, m_{\Delta}$ for the two cases considered here, omitting the $i\epsilon$. The inclusion of the $\Delta$ width also changes the analytic structure. In the following, we denote the configurations corresponding to $\kappa_1, \kappa_2$ with $\Delta$ widths by $\omega_1, \omega_2$. 
\begin{figure}[ht]
\centering
\includegraphics[width=0.5\textwidth]{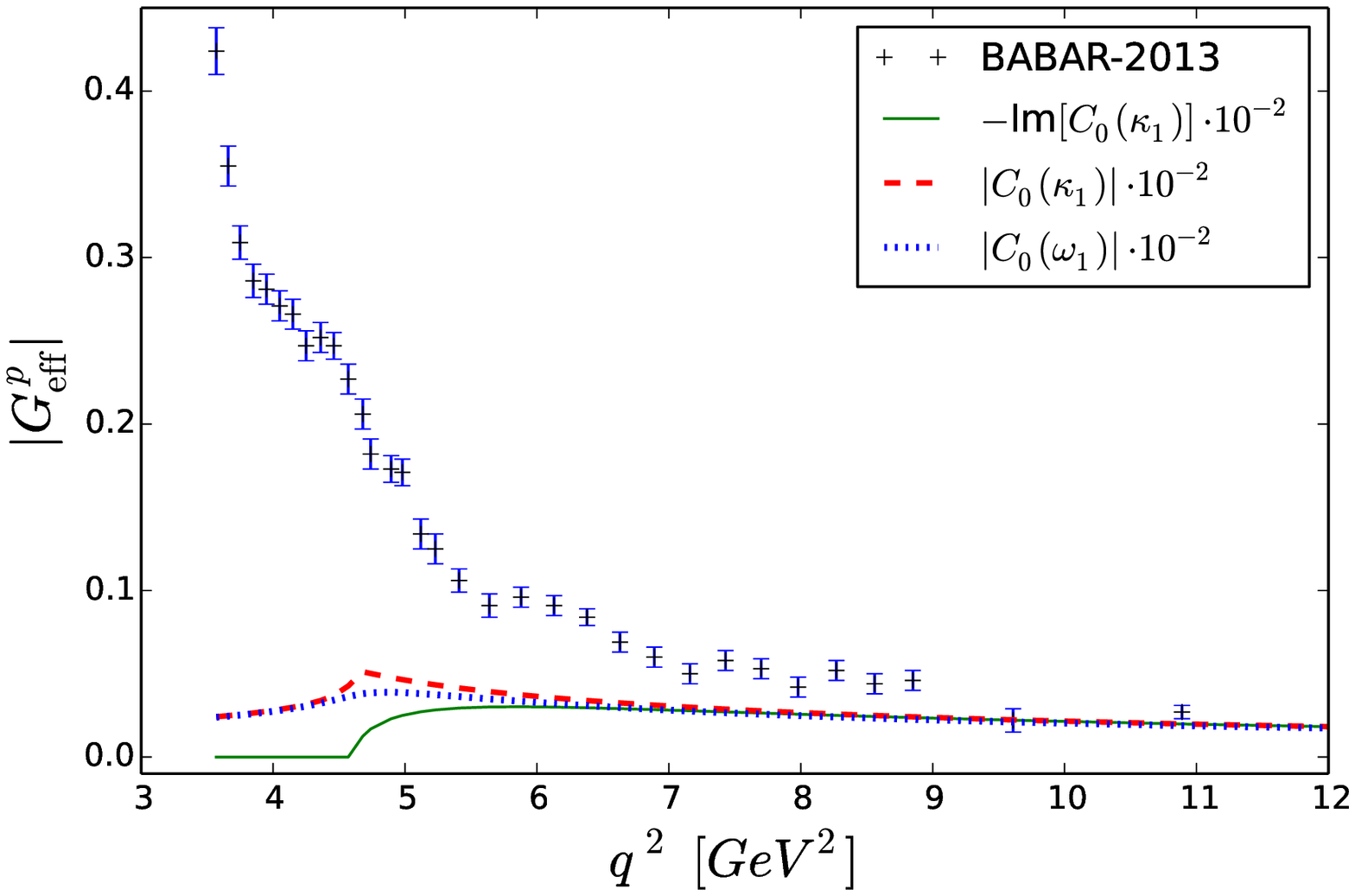}\hglue1mm
\includegraphics[width=0.5\textwidth]{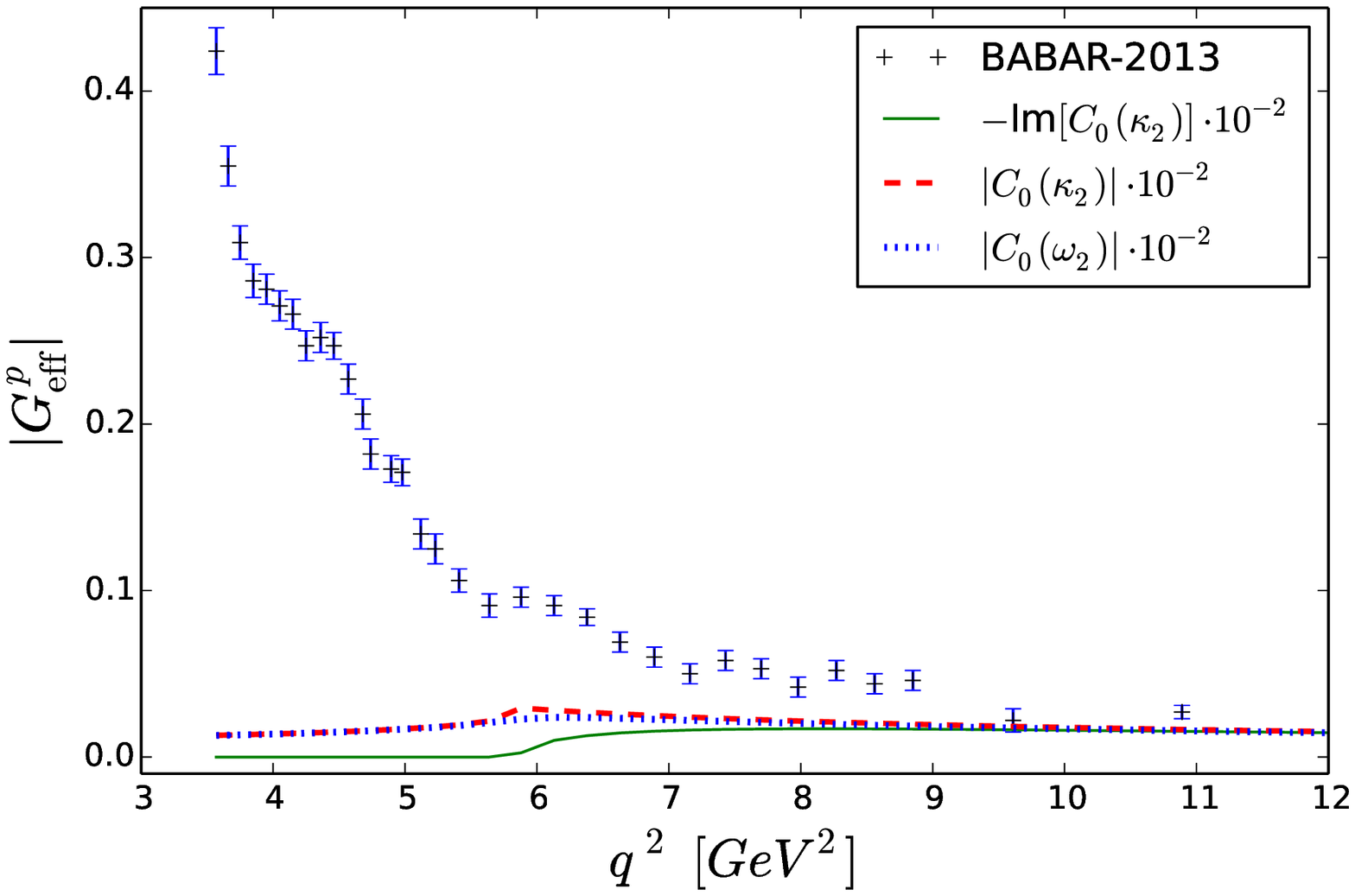}
\caption{The $q^2$-dependence from the scalar Passarino-Veltman triangle diagrams with virtual $N\bar{\Delta}\pi$ and $\Delta\bar{\Delta}\pi$
states compared to $G_{\rm eff}^p$  from Ref.~\cite{Lees:2013ebn}.}\label{fig:pv}
\end{figure}
We show the absolute value and imaginary part of the configurations $\kappa_1$ and $\kappa_2$ in Fig.~\ref{fig:pv}. Also shown is how the inclusion of the $\Delta$ width partly smears out the cusp effect. Taking the loop momenta in the numerator into account, one can reduce the graph to a sum of $n$-point functions with $n\leq 3$. The momenta only partly cancel against those in the poorly known form factors, so that an additional smearing of the result is expected.

\subsection{Cusp fits}
\noindent 
In this section, we show how the scalar parts of the relevant triangle diagram compares to the convex structures in $G_{\rm eff}^p$. Even after inclusion of the width, the remaining enhancements have the right position and shapes to improve a pure pole fit. As in Sec.~\ref{sec:gbw}, we fit only $G_{\rm eff}^p$, include 5 effective pole terms (below threshold) and now replace the explicit resonance terms by the loop structures from the last section. In order to account for the form factors at the vertices and a smearing as discussed in the last section, we include one form factor for each loop
\begin{align}
 F(q^2) = \frac{1}{1 + q^2/\Lambda_{N\Delta/\Delta\Delta}^2}~,
\end{align}
with $\Lambda_{N\Delta/\Delta\Delta}$ the respective fitted cut-off parameter. Additionally, the overall size of the loop contributions is allowed to vary by a fit parameter $f_{N\Delta/\Delta\Delta}$ for each loop. 
\begin{figure}[ht]
\centering
\includegraphics[width=0.6\textwidth]{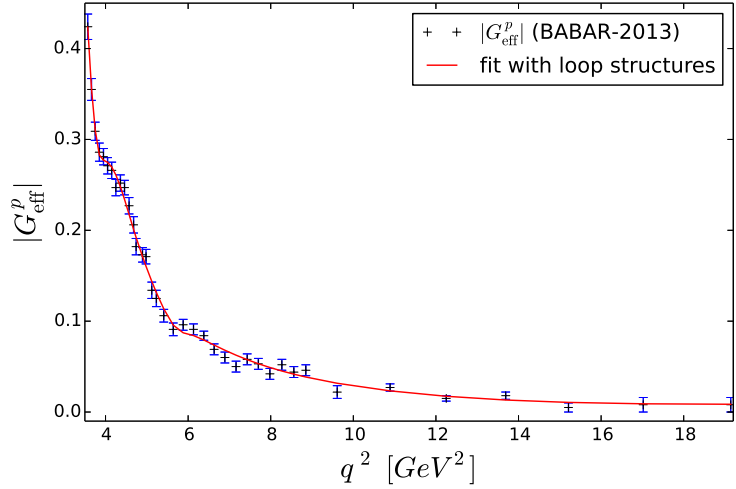}
\caption{Fit including effective pole terms and the scalar parts of the triangle diagrams  with virtual $N\bar{\Delta}\pi$ and $\Delta\bar{\Delta}\pi$ 
states to $G_{\rm eff}^p$ from Ref.~\cite{Lees:2013ebn}.}\label{fig:predloop}
\end{figure}
The fit result with $f_{N\Delta} = 0.02$, $f_{\Delta\Delta} = 0.3$ and $\Lambda_{N\Delta} = 10$ GeV, $\Lambda_{\Delta\Delta} = 1.7$ GeV is shown in Fig.~\ref{fig:predloop}. The fit parameters $f_{N\Delta/\Delta\Delta}$ are of natural size, as expected. 

\section{The unphysical region}\label{sec:unph}
\noindent 
In this section, we discuss the NFFs in the region of $t_0 = 4M_{\pi}^2 < t < t_{ph} = 4m_p^2$ which is not accessible by direct measurements, but by analytic continuation in $t = q^2 = -Q^2$. An additional particle emission from the initial state proton can lower the energy of the (virtual proton) to reach below the threshold, as discussed in Ref.~\cite{Guttmann:2012sq} for the process $p\bar{p}\to e^+e^-\pi_0$. Without model assumptions though, one can relate the information in this region to the physical ones by means of dispersion relations (DRs). In many applications of these, the higher energy parts of the spectral function are not particularly relevant or are suppressed by subtractions. However, it can also be of interest to use (experimental) information from the physical timelike region which only determines the absolute value of the NFFs. To this aim, it is instructive to use a DR for the logarithm, see e.g. Refs.~\cite{Gourdin:1974iq, Geshkenbein:1974gm, Baldini:1998qn, Pacetti:2007zz, Pacetti:2010nv}. In principle, this also allows for a separation of the FF phase $\delta(t)$ and modulus in the representation $G(t) = |G(t)|e^{i\delta(t)}$. The relative phase of $G_E$ and $G_M$ in turn, can be obtained in polarization measurements, as planned for PANDA at FAIR. This phase might help to understand the origin of the structures in $G_{\rm eff}^p$. Moreover, with ideally accurate data in the space- and timelike physical region one could obtain information on both the modulus and phase of the FFs in the unphysical region, including the latter above production threshold.\newline
One can start from a subtracted DR for the function $\ln [G(t)/G(0)]/(t\sqrt{t_0 - t})$. For $t<0$, we evaluate the DR
\begin{align}
 \ln G(t) = \ln G(0) + \frac{t\sqrt{t_0 - t}}{\pi}\int_{t_0}^{\infty}\frac{\ln|G(t')/G(0)|}{t'(t'-t)\sqrt{t'-t_0}}dt' \equiv \int_{t_0}^{\infty} I(t,t_0,t')dt',\label{eq:ln}
\end{align}
where the first term vanishes due to the normalization $G_E(0) = G_M(0)/\mu_p = 1$. Experimental information on this integral equation \eqref{eq:ln} is available in the spacelike region $t<0$ on $G(t)$ and in the timelike region for $t>t_{ph}$ on the modulus $|G(t)|$. One can thus split the integral into the known part above $t>t_{ph}$ and the remaining part with unknown integrand, as suggested in Ref.~\cite{Baldini:1998qn}. The resulting integral equation is commonly denoted as an inhomogeneous Fredholm equation of the first kind \cite{numrecipe}. In general, the solution for the unknown part of $\ln|G(t)|$ can be searched for by discretizing the integral. 
Our first choice would be a number of discretization points equal to the number of input points from the physical region, giving a set of $n$ linear equations with $n$ variables. However, the problem is strongly ill-conditioned, with small changes in the input leading to large changes in the solution. Therefore, additional information is required to solve the original integral equation. We proceed similar to Refs.~\cite{Baldini:1998qn, Pacetti:2007zz} and consider the integral contributions to the logarithm $\ln|G(t)|$ in the spacelike region, using definite values for the known part above $t_{ph}$
\begin{align}
 \ln G(t) - \int_{t_{ph}}^{\infty} I(t,t_0,t')dt' = \int_{t_0}^{t_{ph}} I(t,t_0,t')dt',\hspace{15pt}t < 0.\label{eq:minint}
\end{align}
\begin{figure}[ht]
\centering
\includegraphics[width=0.5\textwidth]{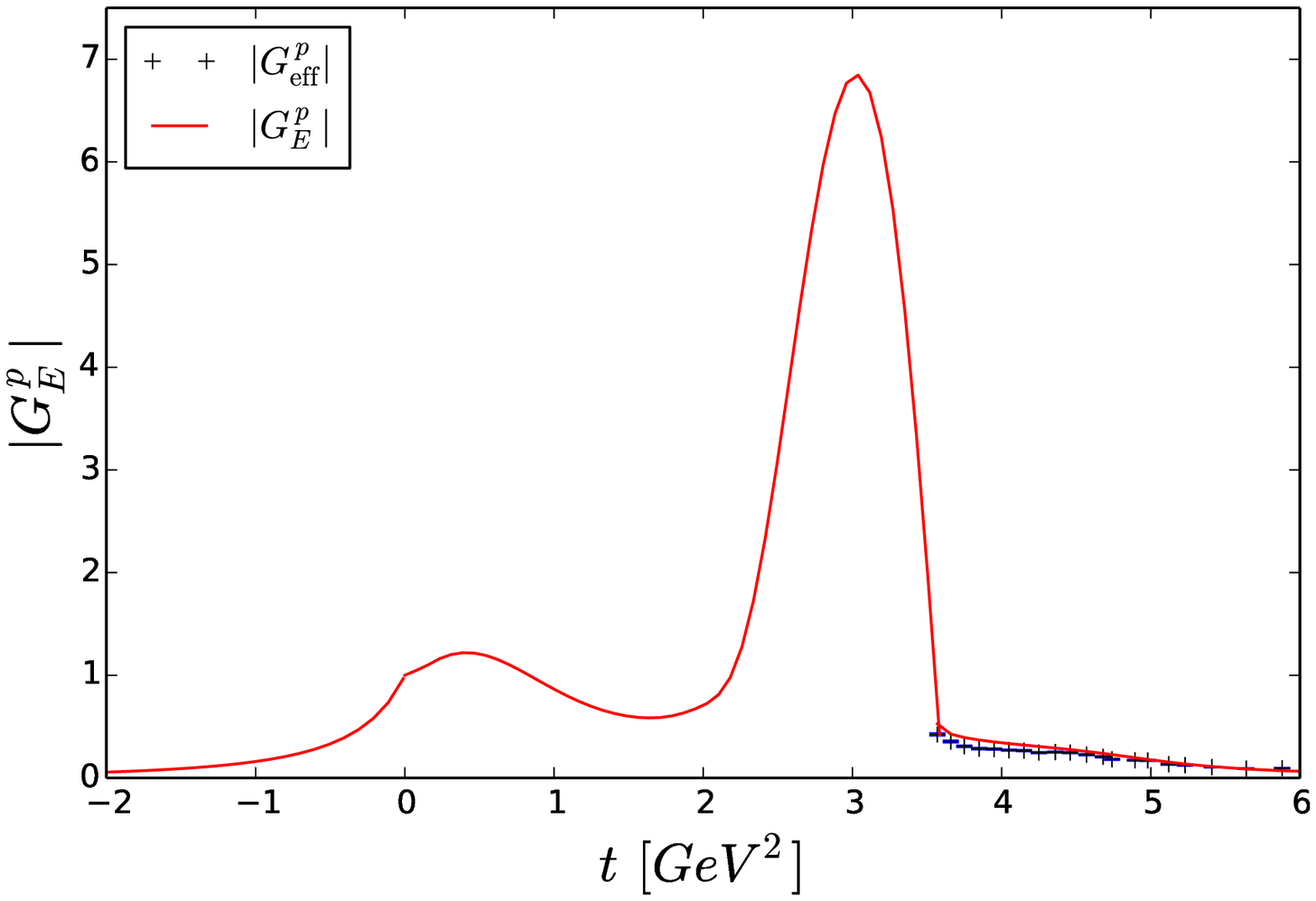}\hglue1mm
\includegraphics[width=0.5\textwidth]{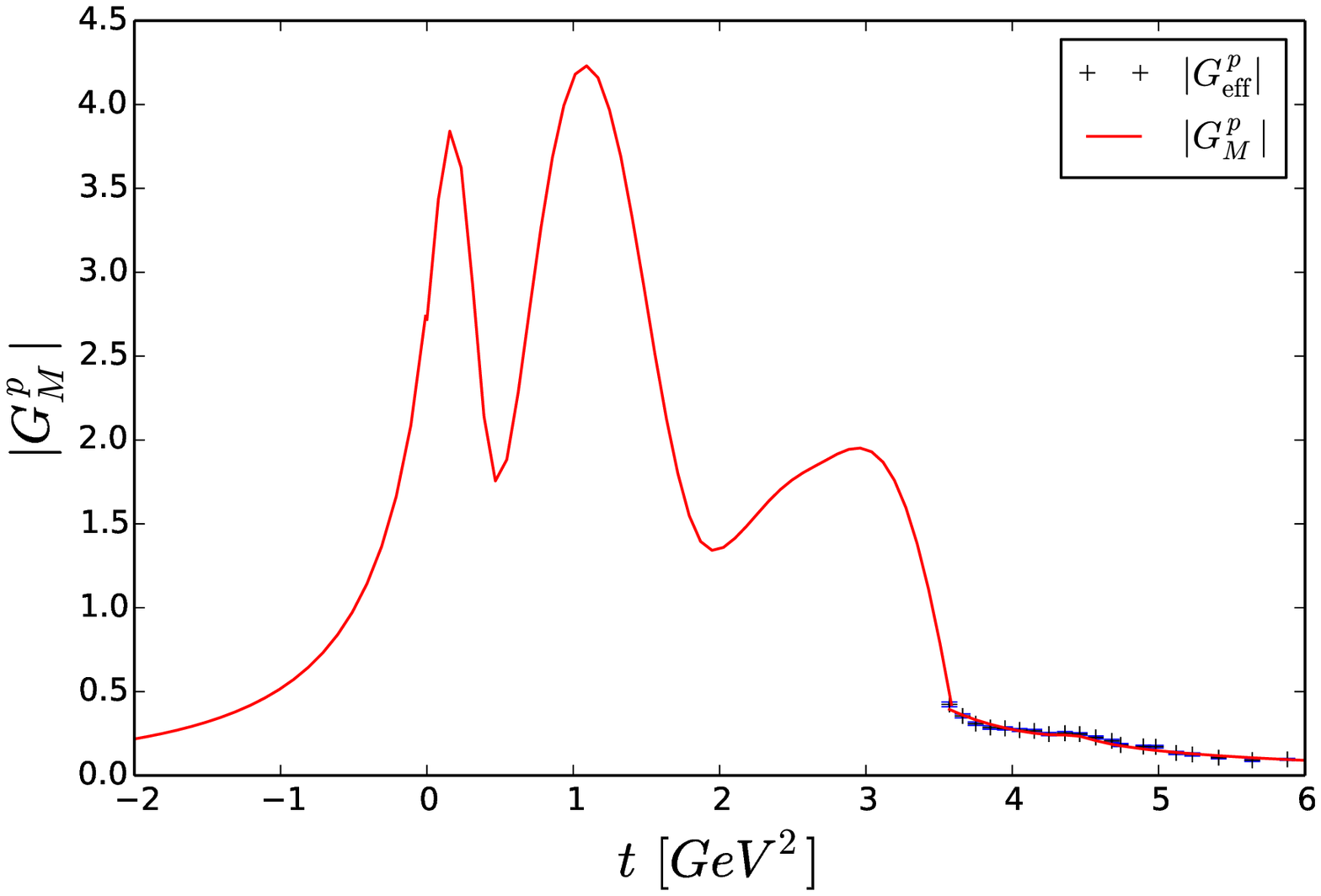}
\caption{An exemplary result for the modulus of electric and magnetic form factor obtained from the logarithmic integral Eq.~\eqref{eq:ln}. The NFFs for $t<0$ are considered via the differential cross sections, therefore the data cannot be shown directly in this form. For $t > t_{ph}$, $|G_M|$ is closer to $|G_{\rm eff}|$ than $|G_E|$, as expected.  We note that large enhancements in the FF modulus just below production threshold are possible and should be evaluated via this method with future PANDA data.}\label{fig:ln}
\end{figure}
In contrast to Refs.~\cite{Baldini:1998qn, Pacetti:2007zz}, we use as input for the lhs of Eq.~\eqref{eq:minint} our discretized result of a simultaneous fit to data in all accessible regions, see Sec.~\ref{sec:phi}, and obtain as an example fit the result shown in Fig.~\ref{fig:ln}. We minimize the difference between left- and right-hand-side  of Eq.~\eqref{eq:minint}, while also limiting the total curvature of the FFs in the unphysical region, $r = \int_{t_0}^{t_{ph}} ((d^2|G(t)|/dt^2))^2dt$. For the result, we find a large dependence on the strength of the curvature limitation and on the range that we use for input from $t < 0$. Thus reliable errors can not be quantified here. 

However, we want to point out that large enhancements in the FF modulus just below production threshold are possible.
In the example shown in Fig.~\ref{fig:ln}, one can see such an enhancement just below $t_{ph}$, as expected in the case of a baryonium pole. For the planned precision of the forthcoming measurements at PANDA, these possibilities should be further evaluated. Encouraging in this regard are also previous results for the pion FF \cite{Baldini:1998qn}, where the predictive power of this method can be impressively illustrated. A major source of complication in the nucleon case is the neccessity of two FFs and their separation. The emphasis on measuring  the angular distribution at PANDA will have particular impact on this separation.

\section{Discussion and Conclusions}\label{sec:disc}
\noindent 
In this paper, we have examined the em $p\bar{p}$ creation and the mechanisms that dominate it in the domain of nonperturbative QCD. Specifically, we have discussed possible contributions to the NFFs corresponding to a vector meson recently listed in the PDG as $\phi(2170)$ or from FSI at the $N\bar{\Delta}$ + c.c. and $\Delta\bar{\Delta}$ thresholds.\newline
We have included the $\phi(2170)$ in simultaneous fits to proton and neutron FFs for space- and timelike momenta and found good agreement with the existing data. In particular, we included recent measurements on the neutron effective FF. In contrast to the previous FENICE experiment and analyses of this, the recent SND data shows a very similar behavior to the proton effective FF over a large range, which we can describe well in our approach. However, the range around the $\phi(2170)$ calls for further neutron measurements to allow for a determination of the isospin channel of the structures in $G_{\rm eff}^p$.\newline
It may be worthwhile mentioning here, that similar fits to data only in the spacelike region, as performed in Ref.~\cite{Lorenz:2012tm}, found exactly two (``effective'') poles in the physical timelike region, one at $2.14$~GeV and one at $2.4$~GeV, each $10-20$~MeV below the $N\bar{\Delta}$ and $\Delta\bar{\Delta}$ thresholds, respectively. Accordingly, we have also examined possible contributions from the final state interactions at these thresholds. Taking approximations for the FSI into account allows for a similarly good description of the $G_{\rm eff}^p$ data as the inclusion of the $\phi(2170)$. The occurrence of peaks in $G_{\rm eff}^p$ around both $\Delta$ thresholds might favor this explanation. However, future calculations should include the singularity structures of any possibly contributing diagram and all interferences. In particular for the case that some structures indeed exist below the thresholds, a resummation of the FSI diagrams is clearly necessary to calculate the pole of such a bound state. In this context it may be of interest, that the small deviation at $\sim 2.25$~GeV lies close to the $\Lambda\bar{\Lambda}$ threshold.\newline
In order to distinguish between the possible effects we are awaiting polarization measurements at FAIR from which one can extract the relative phase of $G_E$ and $G_M$. This will also improve the precision of the analytic continuation to the region of a possible baryonium pole. Using logarithmic dispersion relations, we have found that such a pole with large contributions to the NFFs could exist. 

\section*{Acknowledgements}

We thank Johann Haidenbauer and Christoph Hanhart for useful comments.
This work is supported in part by Deutsche Forschungsgemeinschaft (Sino-German CRC 110) and by the
Helmholtz Association under contract HA216/EMMI.


\end{document}